\documentclass[twocolumn, aps,prl,superscriptaddress,amsfonts]{revtex4-1}
\usepackage{amsmath}
\usepackage{graphicx}
\usepackage{bm}
\usepackage{array}
\usepackage{dcolumn}
\usepackage{hepparticles}
\usepackage{heppennames}
\usepackage{hepnicenames}
\usepackage{epstopdf}

\begin{document}

\title{Experimental Comparison of Two Quantum Computing Architectures}

\author{N. M. Linke}
\affiliation{Joint Quantum Institute and Department of Physics, University of Maryland, College Park, MD 20742}
\author{D. Maslov} 
\affiliation{National Science Foundation, Arlington, VA  22230}
\affiliation{Joint Center for Quantum Information and Computer Science, University of Maryland, College Park, MD  20742}
\author{M. Roetteler}
\affiliation{Microsoft Research, Redmond, WA  98052}
\author{S. Debnath}
\affiliation{Joint Quantum Institute and Department of Physics, University of Maryland, College Park, MD 20742}
\author{C. Figgatt}
\affiliation{Joint Quantum Institute and Department of Physics, University of Maryland, College Park, MD 20742}
\author{K. A. Landsman}
\affiliation{Joint Quantum Institute and Department of Physics, University of Maryland, College Park, MD 20742}
\author{K. Wright}
\affiliation{Joint Quantum Institute and Department of Physics, University of Maryland, College Park, MD 20742}
\author{C. Monroe}
\affiliation{Joint Quantum Institute and Department of Physics, University of Maryland, College Park, MD 20742}
\affiliation{Joint Center for Quantum Information and Computer Science, University of Maryland, College Park, MD  20742}
\affiliation{IonQ, Inc., College Park, MD  20742}

\begin{abstract}
We run a selection of algorithms on two state-of-the-art 5-qubit quantum computers that are based on different technology platforms. One is a publicly accessible superconducting transmon device \cite{IBM} with limited connectivity, and the other is a fully connected trapped-ion system \cite{Debnath16}. Even though the two systems have different native quantum interactions, both can be programmed in a way that is blind to the underlying hardware, thus allowing the first comparison of identical quantum algorithms between different physical systems. We show that quantum algorithms and circuits that employ more connectivity clearly benefit from a better connected system of qubits.  While the quantum systems here are not yet large enough to eclipse classical computers, this experiment exposes critical factors of scaling quantum computers, such as qubit connectivity and gate expressivity.  In addition, the results suggest that co-designing particular quantum applications with the hardware itself will be paramount in successfully using quantum computers in the future.
\end{abstract}

\maketitle

Inspired by the vast computing power a universal quantum computer could offer, several candidate systems are being explored. They have allowed experimental demonstrations of quantum gates, operations, and algorithms of ever increasing sophistication. Recently, two architectures, superconducting transmon qubits \cite{Barends14, Corcoles2015, Riste2015, Ofek16, Takita16} and trapped ions \cite{Monz16, Debnath16}, have reached a new level of maturity. They have become fully programmable multi-qubit machines that provide the user with the flexibility to implement arbitrary quantum circuits from a high-level interface. This makes it possible for the first time to test quantum computers irrespective of their particular physical implementation. 

While the quantum computers considered here are still small scale and their capabilities do not currently reach beyond small demonstration algorithms, this line of inquiry can still provide useful insights into the performance of existing systems and the role of architecture in quantum computer design. These will be crucial for the realization of more advanced future incarnations of the present technologies.

The standard abstract model of quantum computation assumes that interactions between arbitrary pairs of qubits are available. However, physical architectures will in general have certain constraints on qubit connectivity, such as nearest-neighbor couplings only. These restrictions do not in principle limit the ability to perform arbitrary computations, since SWAP operations may be used to effect gates between arbitrary qubits using the connections available. For a general circuit, reducing a fully-connected system to the more sparse star-shaped or linear nearest-neighbor connectivity requires an increase in the number of gates of $O(n)$, where $n$ is the number of qubits \cite{cheung2007translation}. How much overhead is incurred in practice depends on the connections used in a particular circuit and how efficiently they can be matched to the physical qubit-to-qubit interaction graph.

\begin{figure}[t]
\includegraphics[width=0.99\columnwidth]{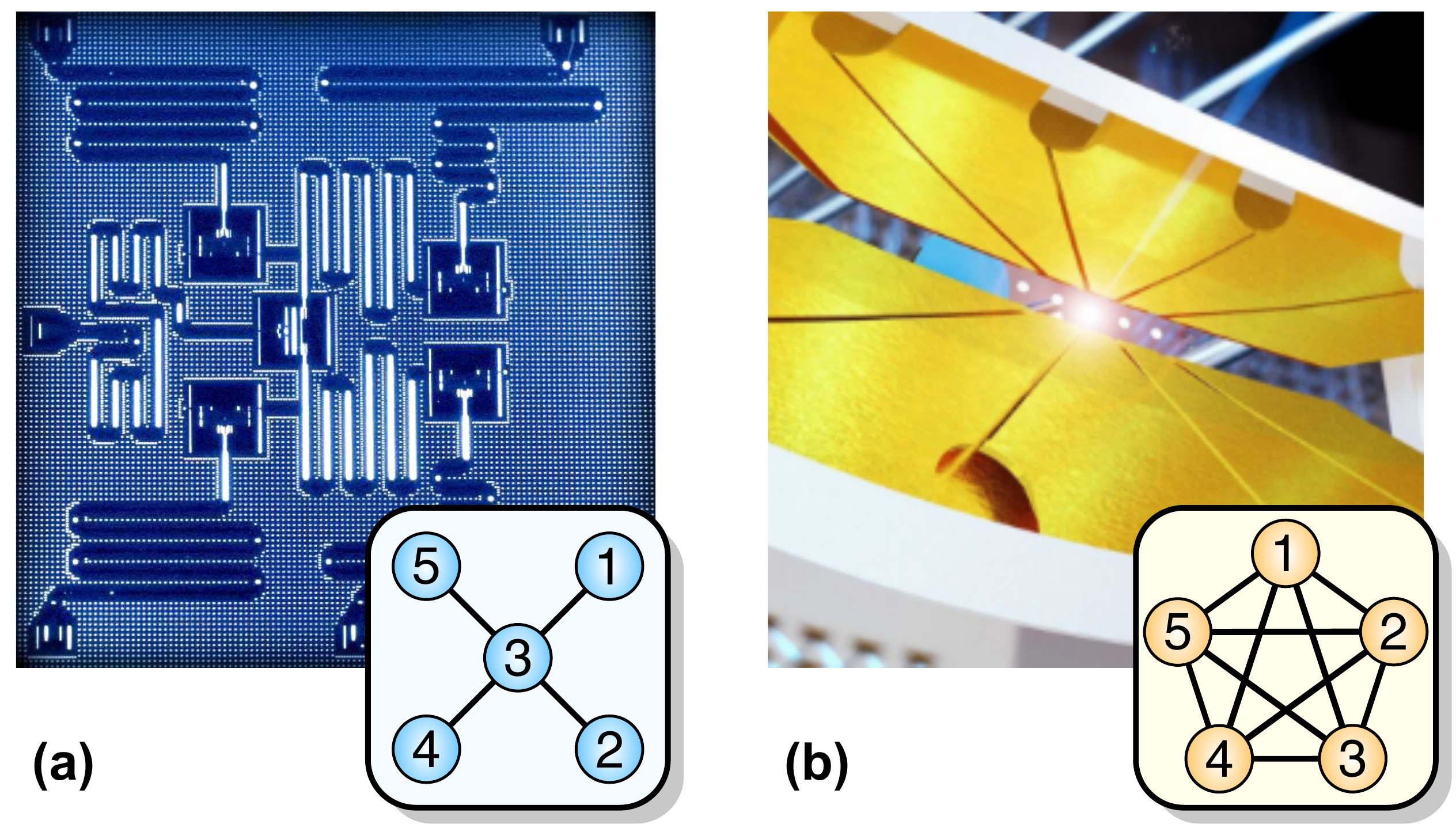}
\caption{Graphic representations of the two systems: (a) the superconducting qubits connected by microwave resonators (Credit: IBM Research), and (b) the linear chain of trapped ions connected by laser-mediated interactions. Insets: Qubit connectivity graphs, (a) star-shaped and (b) fully connected.}
\label{fig:fig1}
\end{figure}

In this article, we make use of the public access recently granted by IBM to a $5$-qubit superconducting device (illustrated in fig.\ref{fig:fig1}(a)) via their ``Quantum Experience'' cloud service \cite{IBM}. This allows us to repeat algorithms that we perform in our own ion trap experiment on an independent quantum computer of identical size and comparable capability but with a different physical implementation at its core.

\section*{Physical Systems} 
The ion trap system consists of five $^{171}$Yb$^+$ ions which are confined in a linear Paul trap and laser-cooled close to their motional ground state (see fig.\ref{fig:fig1}(b)) \cite{Debnath16}. The qubits are magnetic field-insensitive pairs of states in the hyperfine-split $^2S_{1/2}$ ground-level of each atom, which gives a qubit frequency of $12.642821\;\textrm{GHz}$. All control and measurement is performed optically. State preparation and readout are accomplished by optical pumping and state-dependent fluorescence detection \cite{Olmschenk07}. Qubit operations are realized via pairs of Raman beams, derived from a single $355\:$nm mode-locked laser. These optical controllers consist of an array of individual addressing beams and a counter-propagating global beam that illuminates the entire chain \cite{Debnath16}. Single-qubit rotations are driven by a Raman beat-note of defined amplitude, phase, and duration resonant with the qubit frequency. Two-qubit operations are produced by applying Raman beams to a pair of ions, with beat-note frequencies near the motional sidebands. This creates an effective XX-Ising interaction between the spins mediated by all modes of motion \cite{Molmer99, Solano99, Milburn00}. We use a pulse-shaping scheme to ensure spin and motion are disentangled at the end of the operation \cite{Zhu06,Choi14}. Since all ions partake in the collective motion of the chain, gates between any pair can be invoked in this way (see inset of fig.\ref{fig:fig1}(b)). The addressing during operations and the distinction between qubits during readout are both achieved by spatially resolving the ions. The fidelities for single- and two-qubit gates are typically $99.1(5)\%$ and $97(1)\%$, respectively. The single-qubit readout fidelity is $99.7(1)\%$ for state $|0\rangle$, and $99.1(1)\%$ for state $|1\rangle$. The latter is lower since off-resonant excitation during readout predominantly causes $|1\rangle\rightarrow|0\rangle$ pumping. The average readout fidelity for an entire $5$-qubit state is $95.7(1)\%$. This is lower than one would expect from the average single-qubit readout fidelity, since there is crosstalk that leads to $|0\rangle\rightarrow|1\rangle$ errors on adjacent channels. Typical gate times are $20\:\mu$s for single- and $250\:\mu$s for two-qubit gates. Spin depolarization is negligible for hyperfine ground level qubits ($T_1 \sim \infty$). The spin-dephasing time ($T_2^*$) is $\sim 0.5\:$s in the current setup, and can be easily extended by suppressing magnetic field noise.

\begin{figure}[t]
\includegraphics[width=0.99\columnwidth]{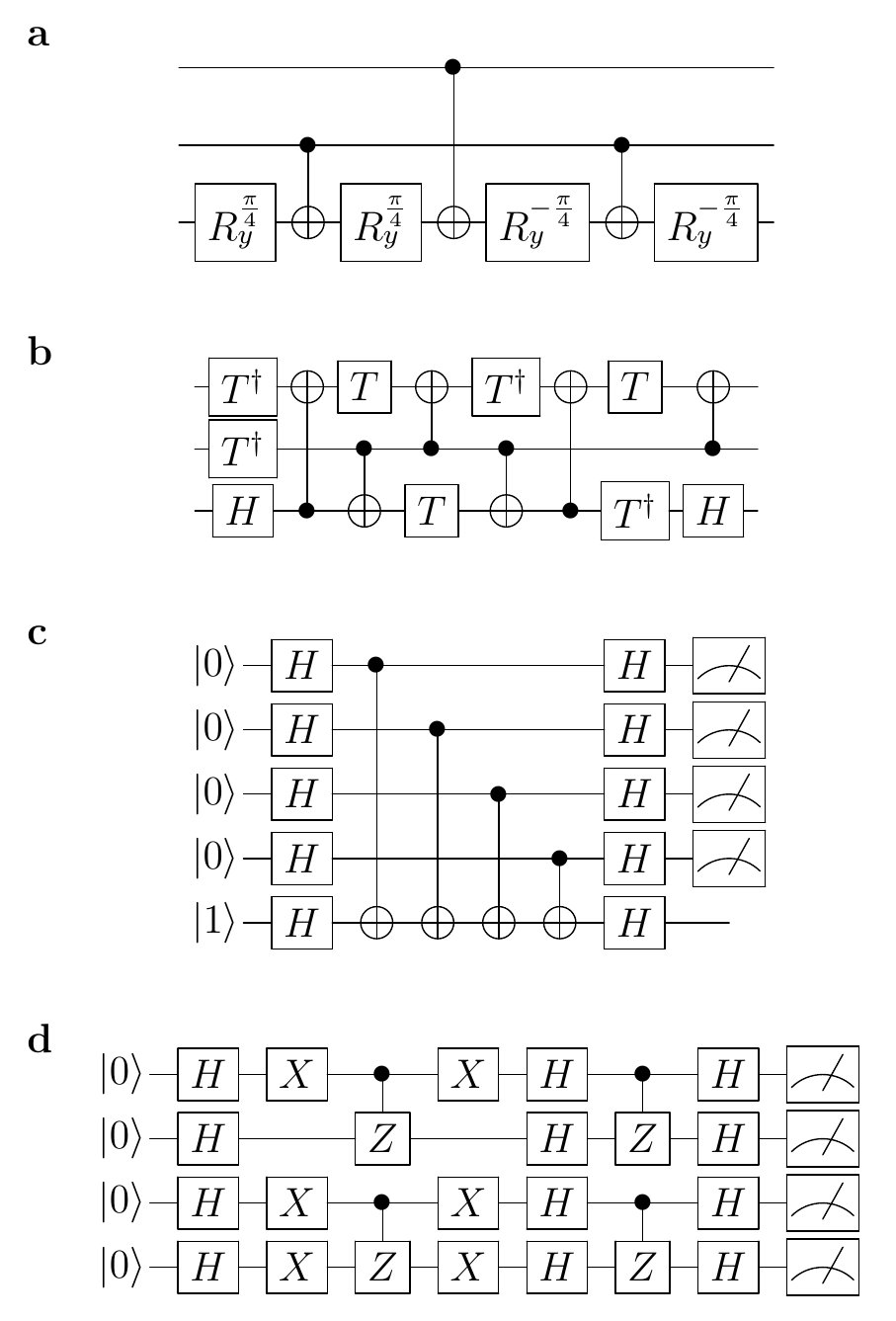}
\caption{High level circuits of the implemented example computations (gates defined in \cite{Nielsen11}): Margolus gate (a), Toffoli gate (b), Bernstein-Vazirani (c), and hidden shift (d). The Bernstein-Vazirani algorithm is shown for the oracle ${\bf c} = (1 1 1 1)$, where all CNOTs are present. The hidden shift diagram represents the shift pattern $s = (1 0 1 1)$, where X-operations are present on qubits $1$, $3$ and $4$.}
\label{fig:circuits}
\end{figure}

In analogy to atoms given by nature, the man-made superconducting circuits in the IBM quantum computer can be thought of as ``artificial atoms'' \cite{devoret13}. They are transmon qubits \cite{Koch2007}, or superconducting islands connected by Josephson junctions and shunt capacitors that provide superpositions of charge states which are insensitive to charge fluctuations. The device used here has a range of qubit frequencies between $5$ and $5.4\:$GHz \cite{IBMpriv}. The qubits are connected to each other and the classical control system by microwave resonators. State preparation \cite{Geerlings13} and readout, as well as single- \cite{Chow10} and two-qubit gates \cite{Chow11}, are achieved by applying tailored microwave signals to this network and measuring the response. Qubits are resolved in the frequency domain during addressing and readout. In the Quantum Experience hardware, the qubits are connected in a star-shaped pattern that provides four two-qubit interactions (see inset fig.\ref{fig:fig1}(a)), which are CNOT gates targeting the central qubit. Single-qubit readout fidelities are typically $\sim96\%$ \cite{IBM}, and the average readout fidelity for an arbitrary $5$-qubit state is $\sim80\%$ \cite{IBMpriv}. Typical gate fidelities are $99.7\%$ and $96.5\%$ for single- and two-qubit gates, respectively. Typical gate times are $130\:$ns for single- and $250-450\:$ns for two-qubit gates, while coherence times are $\sim 60\:\mu$s for both depolarization ($T_1$) and spin dephasing ($T_2$). The publicly accessible system runs autonomously, not requiring any human intervention over many weeks \cite{IBMpriv}. This level of reliability may come at a cost due to drifts between periodic calibrations. Higher connectivity can in general be achieved by coupling $3$-$4$ transmons to one resonator, limited by spectral resolution. The present layout could be modified to provide connections from qubit $1$ to $5$, and $2$ to $4$ \cite{IBMpriv}. Furthermore, other superconducting architectures involving multi-mode resonators \cite{Riste2015} can offer higher connectivity.

On these two machines, we compare a selection of composite gates and algorithms that represent a variety of circuit connectivities. In each case, we map the algorithms to the device by breaking them down into circuits made up of gates native to the specific hardware. We rely on an optimization protocol \cite{Maslov16} to accomplish this task for the trapped ions, and CNOT$+$T/Z$^a$ algebra \cite{Amy13} with further manual optimization to compose the experiments for the IBM machine. The available gate set for the ion trap system consists of the two-qubit XX gate, as well as arbitrary single-qubit R$_\alpha ^\theta$ gate rotations by an angle $\theta$ about any axis (given by $\alpha$) on the equator of the Bloch sphere. We call this the R/XX library. The IBM system makes available the family of gates (X, Y, Z, H, S, CNOT and T \cite{Nielsen11}), known as the Clifford$+$T library. Since each gate is subject to errors, the circuits are optimized to minimize the number of operations used. The resulting gate numbers are optimal for two-qubit gates, and either optimal or close to optimal for single-qubit gates. The total number of single- and two-qubit gates for each algorithm is shown in table \ref{tb:gatecount}. The R/XX library offers a better overall expressive power. However, we note that the Clifford$+T$ library was likely chosen for didactic reasons and is not native to superconducting systems, which do in principle offer continuous parameters for single- and two-qubit gates.  

\begin{table}[t]
\centering
\caption{Single- and two-qubit gate counts for the circuits on the superconducting (star-shaped) and the ion trap (fully connected) system after mapping to the respective hardware using the respective gate libraries. For comparison, the gate counts for a linear nearest-neighbor (LNN) architecture as implemented in \cite{Barends14} are included. We also note the gate count for the Quantum Fourier Transfrom (QFT) for $3$ and $5$ qubits. The latter was implemented in \cite{Debnath16} using a sequence of modular gates that was not optimized for gate count. The QFT-5 cannot be implemented exactly using the current IBM gate library. If we assume Z$^a$ operations are possible, the counts shown as {\textasteriskcentered } are $47$ for single- and $29$ for two-qubit gates.}
\begin{tabular}{|c||r|r||r|r||r|r|}
\hline
\textbf{connectivity} & \multicolumn{2}{c||}{star-shaped} &  \multicolumn{2}{c||}{LNN} & \multicolumn{2}{c|}{fully conn.}\\
\hline
\textbf{hardware} & \multicolumn{2}{c||}{supercond.} &  \multicolumn{2}{c||}{ } & \multicolumn{2}{c|}{ion trap}\\
\hline
\textbf{gate library} & \multicolumn{2}{c||}{Clifford+T} & \multicolumn{2}{c||}{Clifford+Z$^a$} & \multicolumn{2}{c|}{R/XX}\\
\hline
\textbf{gate type}  & single & two & single & two & single & two \\
\hline
\hline
Margolus & 20 & 3 & 20 & 3 & 11 & 3 \\
\hline
Toffoli & 17 & 10 & 9 & 10 & 9 & 5 \\
\hline
Bernstein-Vazirani & 10 & 0-4 & 10 & 0-10 & 14-26 & 0-4 \\
\hline
Hidden Shift & 28-34 & 10 & 20-26 & 4 & 42-50 & 4 \\ 
\hline
\hline
QFT-3 & 42 & 19 & 11 & 7 & 8 & 3\\
\hline
QFT-5 & \textasteriskcentered & \textasteriskcentered & 35 & 28 & 22 & 10\\
\hline
\end{tabular}
\label{tb:gatecount}
\end{table}

In addition to the two systems considered here, the table also gives the numbers for a linear nearest-neighbor (LNN) connectivity architecture as used, e.g., in superconducting qubits \cite{Barends14} as well as semiconductor gated quantum dots \cite{Zajac16}. The numbers in table \ref{tb:gatecount} show that the two-qubit gate count strongly depends on the matching between the circuit and the qubit connectivity graph. The LNN architecture is as efficient as the fully connected system for the hidden shift algorithm, while the star-shaped system incurs overheads; the reverse is true for the Bernstein-Vazirani algorithm (see fig.\ref{fig:circuits}).

\section*{Algorithms}

\subsubsection{Margolus and Toffoli Gate}
\begin{figure}[t]
\includegraphics[width=0.99\columnwidth]{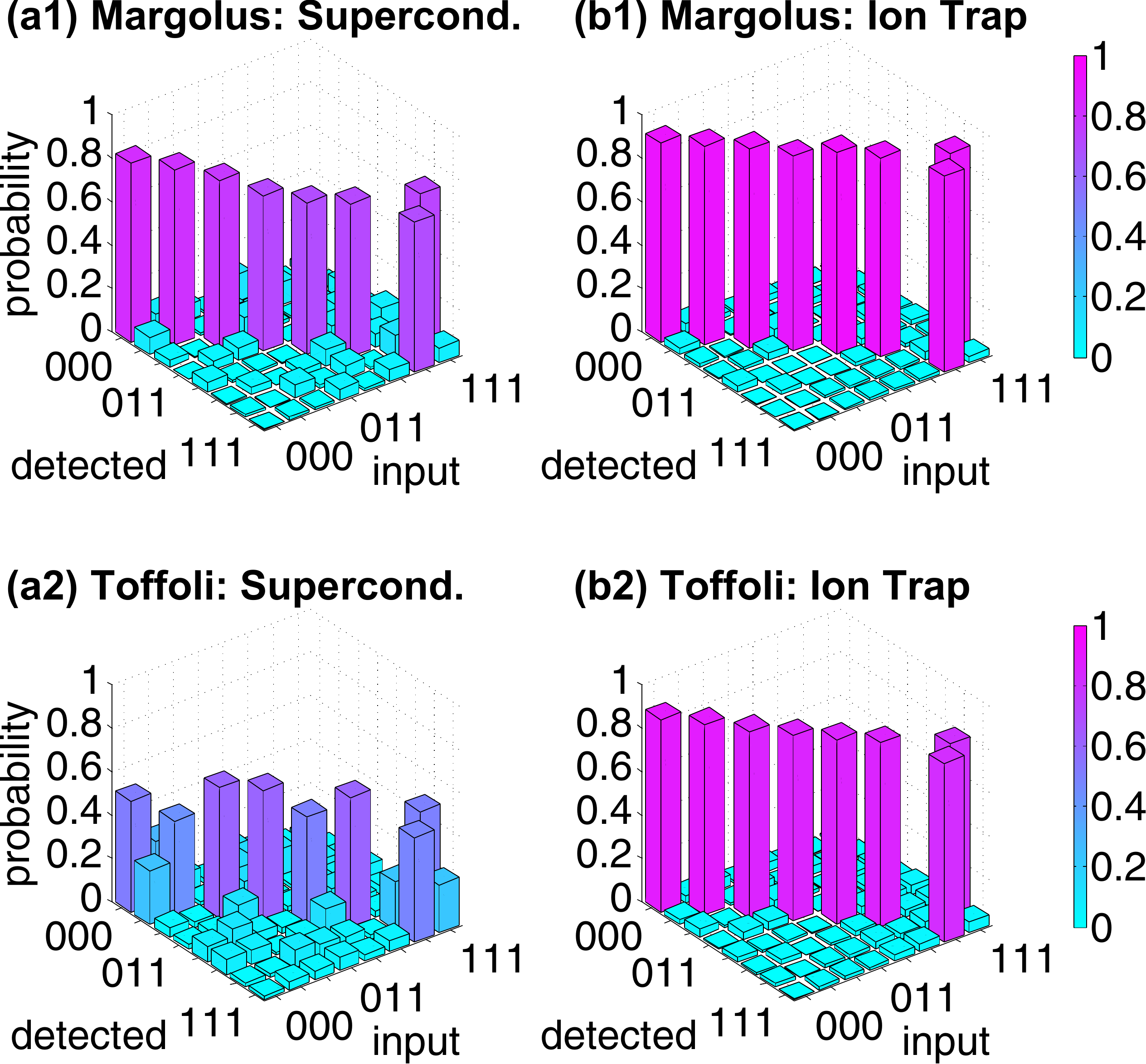}
\caption{Margolus gate results from the star-shaped superconductor (a1) and the fully connected ion trap system (b1). The fidelities are $74.1(7)\%$ and $90.1(2)\%$, respectively. The full Toffoli gate results give success probabilities of $52.6(8)\%$ for the superconducting (a2) and $85.0(2)\%$ for the ion trap (b2) system. The axes represent states as $3$-bit binary numbers. For each input state, the probabilities of detecting each state are shown.}
\label{fig:margolus_toffoli}
\end{figure}

The Toffoli gate is a $3$-qubit controlled-controlled-NOT gate that requires $6$ CNOT gates \cite{Barenco95, Shende09}. It is possible to implement a Toffoli with $5$ entangling gates if the square-root of the CNOT operation is available \cite{Nielsen11}, which is the case with the trapped ion XX gate. The Margolus gate is a simplified version of the Toffoli operation, which introduces an additional phase on the state corresponding to $|{100}\rangle$. It can be realized with just $3$ CNOT gates \cite{divincenzo98, Song04}. The circuits are shown in figure \ref{fig:circuits}(a,b). Note that for the Margolus gate, all entangling operations connect to the same qubit, which means that this circuit can be realized efficiently with star-shaped qubit connectivity. The systems perform this circuit at success probability $74.1(7)\%$ for superconductors and $90.1(2)\%$ for ions (see figure \ref{fig:margolus_toffoli}(a1,b1)). 

The full Toffoli circuit uses the same three qubits as the Margolus implementation so that preparation and measurement errors remain the same. The optimized circuit for the fully connected ion trap system contains $5$ two-qubit gates and the additional operations lower the fidelity to $85.0(2)\%$ (see figure \ref{fig:margolus_toffoli}(b2)). For the star-shaped system, an additional $7$ two-qubit gates are needed to effect the SWAP operations necessary to go from the Margolus to the full Toffoli gate. This leads to a reduced success rate of $52.6(8)\%$ for the superconducting system (\ref{fig:margolus_toffoli}(a2)). Note that the transformation $|a, b, c\rangle \rightarrow |c\oplus ab, b, a\rangle$ may be obtained with the Clifford$+$T library on a star-shaped graph with the provably minimal number of 7 CNOT gates.  We do not consider such input-to-output mappings of the composite gates in this work. However, we always choose the most favorable input-to-output mapping for the IBM star and LNN architectures when executing entire quantum algorithms, which is merely a classical swap between physically measured signals.

\begin{figure*}[t]
\includegraphics[width=0.98\textwidth]{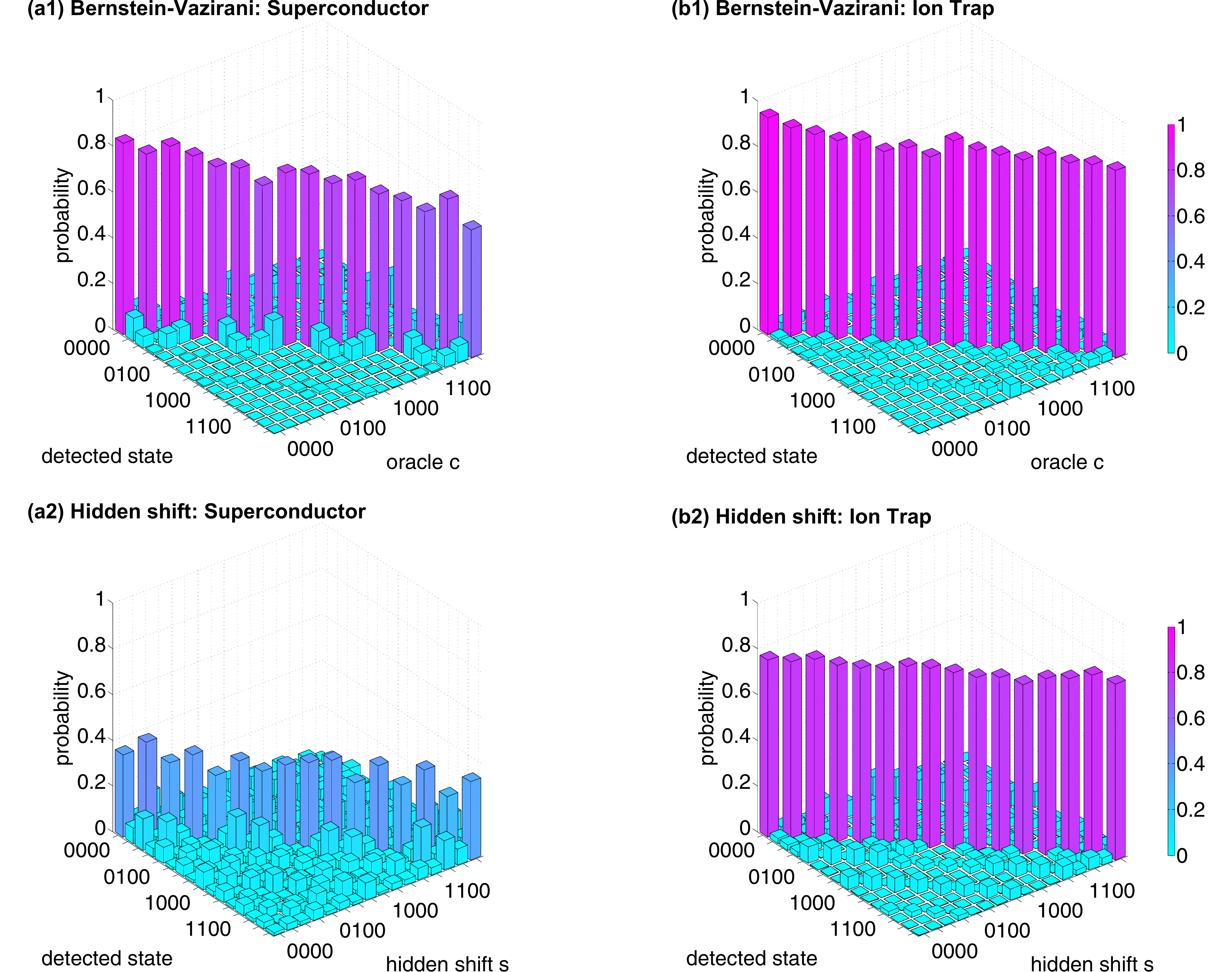}
\caption{Results from the Bernstein-Vazirani algorithm implementing the oracle function $f_{\bf c}({\bf x})=x_0 c_0\oplus c_1 x_1\oplus c_2 x_2\oplus c_3 x_3$ for all possible $4$-bit oracles \textbf{c} performed on the star-shaped (a1) and the fully connected (b1) systems. The average success probabilities are $72.8(5)\%$ for the superconductor and $85.1(1)\%$ for the ion trap system. Hidden shift algorithm for $f({\bf x})=x_0 x_1\oplus x_2 x_3$. All possible $4$-bit shifted oracle functions are implemented on the superconducting system (a2) as well as the ion trap (b2). The average success probabilities are $35.1(6)\%$ and $77.1(2)\%$, respectively. The axes represent states and oracle parameters as $4$-bit binary numbers.}
\label{fig:hiddenshift}
\end{figure*}

\subsubsection{Bernstein-Vazirani and Hidden Shift Algorithms}
In the Bernstein-Vazirani algorithm, an oracle implements the function $f_{\bf c}({\bf x})={\bf x} \cdot {\bf c}$. The algorithm finds the unknown bit string \textbf{c} in a single shot. In the oracle, \textbf{c} is encoded in a pattern of CNOT gates, all of which target the ancilla qubit \cite{Bernstein97}. As can be seen from the circuit in figure \ref{fig:circuits}(c), the entire algorithm maps well onto a star-shaped architecture. This algorithm is very similar to a parity check circuit used in error correction applications, and indeed the IBM system was laid out with this application in mind \cite{Takita16}. The single-shot success probabilities are $72.8(5)\%$ for the star-shaped superconducting system and $85.1(1)\%$ for the fully connected ion trap system (figure \ref{fig:hiddenshift}(a1,b1)).

To compare this to a similar algorithm with different connectivity requirements, we implement the hidden shift algorithm \cite{Dam02} for a black box bent function \cite{Childs13, Roetteler10}. An oracle implements the shifted version $f($\textbf{x}$+$\textbf{s}$)$ of the known Boolean function $f$. We want to determine the $n$-bit string \textbf{s} that constitutes the ``hidden shift''. For a subset of Boolean functions, there exists a quantum algorithm that can solve this problem in a single oracle query, while classical algorithms require $\Omega(\sqrt{2^n})$ queries. This subset contains functions which have a flat Fourier spectrum and whose dual $f^\sim$ can be calculated efficiently, i.e. so-called bent functions of the Maiorana-McFarland class \cite{Roetteler10}. Here we choose the $4$-bit function $f($\textbf{x}$)=x_1 x_2\oplus x_3 x_4$ for which $f=f^\sim$. We implement all possible $4$-bit shift patterns \textbf{s} using the circuit shown in figure \ref{fig:circuits}(d). The algorithm output state directly corresponds to the hidden shift \textbf{s}. The circuit involves gates between two disconnected pairs of qubits, which creates an overhead of $6$ two-qubit gates for a star-shaped architecture. The results are shown in figure \ref{fig:hiddenshift}(a2,b2). The fidelity of the fully connected ion trap implementation is $77.1(2)\%$, compared to $35.1(6)\%$ for the superconducting device. The numerical values of the data plotted in figure \ref{fig:margolus_toffoli} and \ref{fig:hiddenshift} are reproduced as tables in the Appendix.

The errors in both devices appear concentrated in certain sets of states, leading to patterns in the off-diagonal elements of the result plots (see figure \ref{fig:hiddenshift}). These highly structured signatures suggest that systematic errors dominate, especially readout errors. The grouped patterns such as in figure \ref{fig:hiddenshift}(a1) indicate flips of the least-significant bits, while parallel lines correspond to the most significant bits changing their state. In the trapped ion results, these lines can be modulated in height due to read-out crosstalk and are more pronounced on the lower-numbered state side due to $1\rightarrow 0$ being the dominant detection error channel. Finally, we stress that comparing quantum computations across systems depends on the specifics of error propagation, which will vary between different hardware implementations, through their particular connectivity and physical errors. We summarize the success probabilities for the implemented circuits on both machines in Table \ref{tb:fidelities}. We also show the expected values for two simple error propagation models based on the errors of the individual gates $\epsilon_g$ and of $M$-qubit single-shot readout $\epsilon_M$ for both systems. The first model assumes random error propagation per operation with overall error $(1-\epsilon_M)^M (1-\epsilon_g)^{\sqrt{N}}$, where $N$ is the number of gates. Since the errors for each step are independent and comparable to a random walk, the overall error involves $\sqrt{N}$ factors. The second model is based on systematic (coherent) over- or under-rotations with overall error $(1-\epsilon_M)^M (1-\epsilon_g)^N$, which accumulates with $N$ factors. We see that the numbers are broadly consistent, with systematic errors better predicting the superconducting system while the ion trap performance falls in between the two. The superconducting Hidden Shift algorithm is the only example with a significantly lower experimental result, perhaps from inhomogeneous errors in the device.

\section*{Outlook}
Comparing quantum computing architectures involves many interrelated factors.  Quantum gate operation fidelities, qubit numbers, primitive gate speeds, and coherence times are obviously important low-level metrics in a large scale quantum computer.  The results presented here show higher absolute fidelities and coherence times in the trapped ion system, with higher clock speeds for the superconducting system.  However, these metrics are moving targets: while these systems are the most advanced and versatile quantum computing platforms built to date, both technologies are currently advancing rapidly.  

In any case, such metrics should not be considered in isolation.  Our comparison points to important higher level considerations in scaling a quantum computer.  The overall performance of a quantum circuit and the ``time to solution'' will depend critically on architectural restrictions, qubit connectivity, gate reconfigurability, and gate expressivity, and these attributes will become ever more important as the system is scaled up.  Even with 5-qubit systems, we find that the qubit connectivity graph is best co-designed to mirror the structure of the particular quantum circuit and that the choice of a more expressive gate library affects the efficiency of the computations. 

The physical scaling of each of these leading technologies has many challenges, and how they will be connected and reconfigured at large scales is an open question.  One of the biggest challenges is the management of the control complexity in larger systems and potential cross-talk from overlapping qubit interactions or control buses.  In most superconducting designs, there are many current-carrying wires necessary for control and biasing the individual qubits, and this may be difficult to route through a large superconducting chip \cite{Barends14, Corcoles2015, Riste2015, Ofek16, Takita16}.  It will likely become a great challenge to manage the dilution refrigerator heat budget with such circuitry.  Alternative modular superconducting architectures improve connectivity by integrating qubits with microwave cavity modes, at the expense of significant added volume per qubit \cite{Brecht2016}.  Ion trap designs will hinge upon the stable and accurate delivery of laser beams (or near-field microwave sources) to address each qubit individually in a vacuum chamber.  The fully connected nature of the ion trap architecture may not scale to arbitrarily large numbers of qubits, owing to the spectral overlap of collective normal modes of motion. 
However, full connectivity between $20-100$ trapped ion qubits appears possible \cite{Debnath16} and a modular approach for scaling to much larger systems with high connectivity and distance-independent operations seems promising \cite{Monroe14, Kielpinski02}.  In any hardware, an automated calibration procedure and powerful user interface will likely provide a higher level of integration. Such system-level attributes will become even more important as quantum circuits grow in complexity, regardless of physical platform.

\section*{acknowledgements}
We thank D. L. Moehring, J. Kim, and K. R. Brown for key discussions, and J. Gambetta and J. Chow at IBM for their assistance in interfacing with the IBM Quantum Experience project. This work was supported by the ARO with funds from the IARPA LogiQ program, the AFOSR MURI program on Optimal Quantum Circuits, and the NSF Physics Frontier Center at JQI. DM acknowledges support by the NSF. Any opinion, finding, and conclusions or recommendations expressed in this material are those of the authors and do not necessarily reflect the views of the NSF, IBM, or any of their employees.

\begin{table}[t]
\centering
\caption{Summary of the achieved success probabilities for the implemented circuits, in percent. The observed probabilities (``obs") are tabulated alongside two simple error propagation models given the gate number $N$ and the individual gate and readout errors of the two systems encapsulated in the parameters $\epsilon_g$ and $\epsilon_M$, respectively (see main text). The first estimate  assumes random (``rand") error propagation with overall error $(1-\epsilon_g)^{\sqrt{N}}$ while the second is based on systematic (``sys") coherent over- or under-rotations with overall error $(1-\epsilon_g)^N$, where $N$ is the number of gates. The readout error for $M$ qubits is $(1-\epsilon_M)^M$ in both cases.}
\begin{tabular}{|c||c|r|r||c|r|r|}
\hline
\textbf{connectivity} & \multicolumn{3}{c||}{star-shaped} & \multicolumn{3}{c|}{fully conn.}\\
\hline
\textbf{hardware}  & \multicolumn{3}{c||}{supercond.} &  \multicolumn{3}{c|}{ion trap}\\
\hline
\textbf{success prob}/\% \hspace{0.1mm}& obs & rand & sys & obs & rand & sys \\
\hline
\hline
Margolus & 74.1(7) & 82 & 75 & 90.1(2) & 91 & 81 \\
\hline
Toffoli & 52.6(8) & 78 & 59 & 85.0(2) & 89 & 78 \\
\hline
Bernstein-Vazirani & 72.8(5) & 80 & 74 & 85.1(1) & 90 & 77 \\
\hline
Hidden Shift & 35.1(6) & 75 & 52 & 77.1(2) & 86 & 57 \\ 
\hline
\end{tabular}
\label{tb:fidelities}
\end{table}

\section*{appendix}

The detailed results from the algorithms presented in Figs. 3-4 are shown below as tables containing numeric probabilities. The target populations, with nominal unit probabilities, are highlighted in yellow. The others, representing errors, show a bar graph scaled from 0 to 0.1 to emphasize the systematic error patterns.

\begin{figure*}[b]
\includegraphics[width=0.8\textwidth]{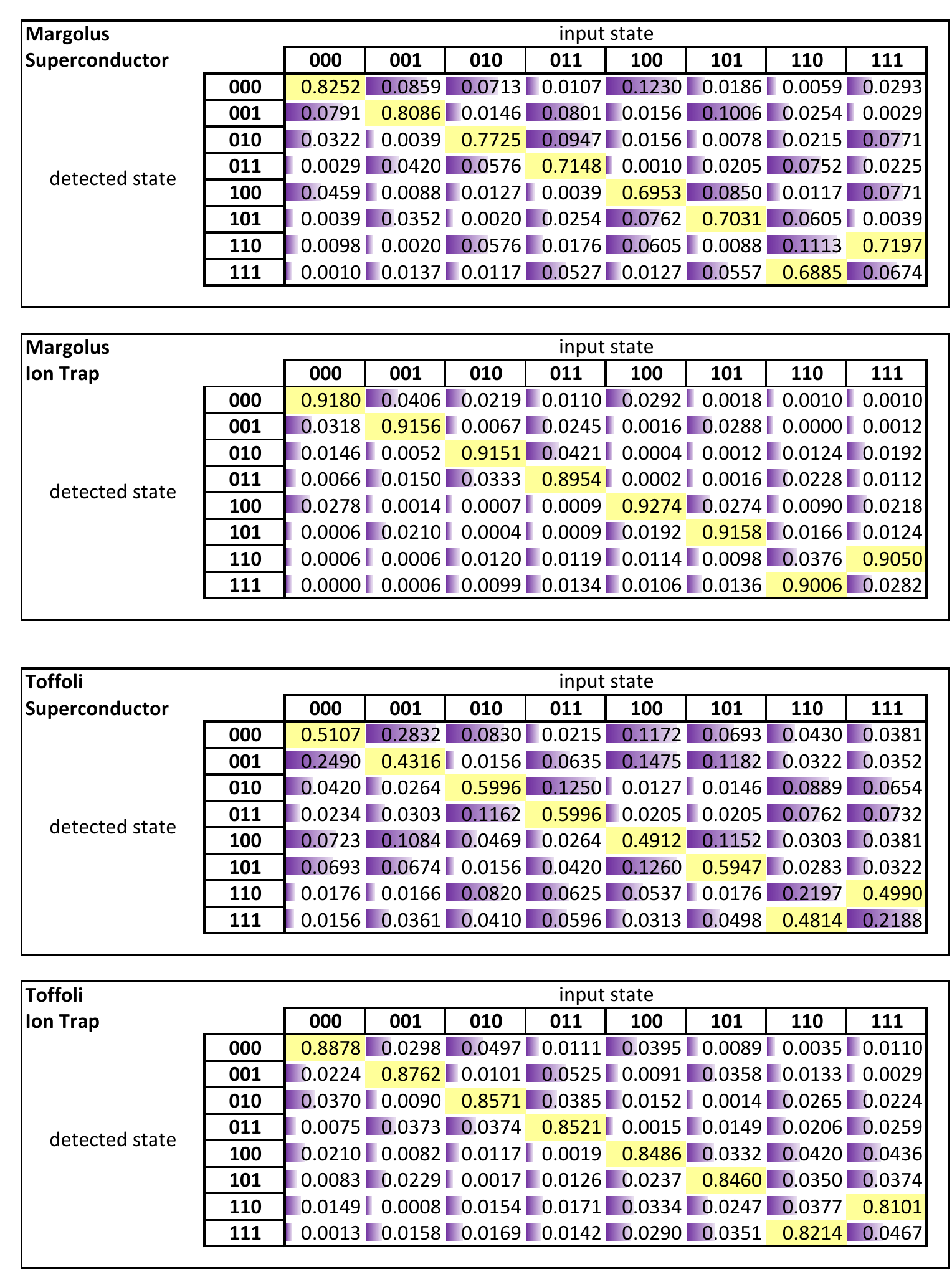}
\caption{Numerical quantum computer 3-qubit input/output matrix for the Margolis gate (top two panels) and Toffoli gate (bottom two panels), corresponding to Fig. 3 of the main text.  For each gate, the results from both superconductor and ion trap quantum computers are displayed.}
\end{figure*}

\begin{figure*}[b]
\includegraphics[height=0.9\textwidth, angle=90]{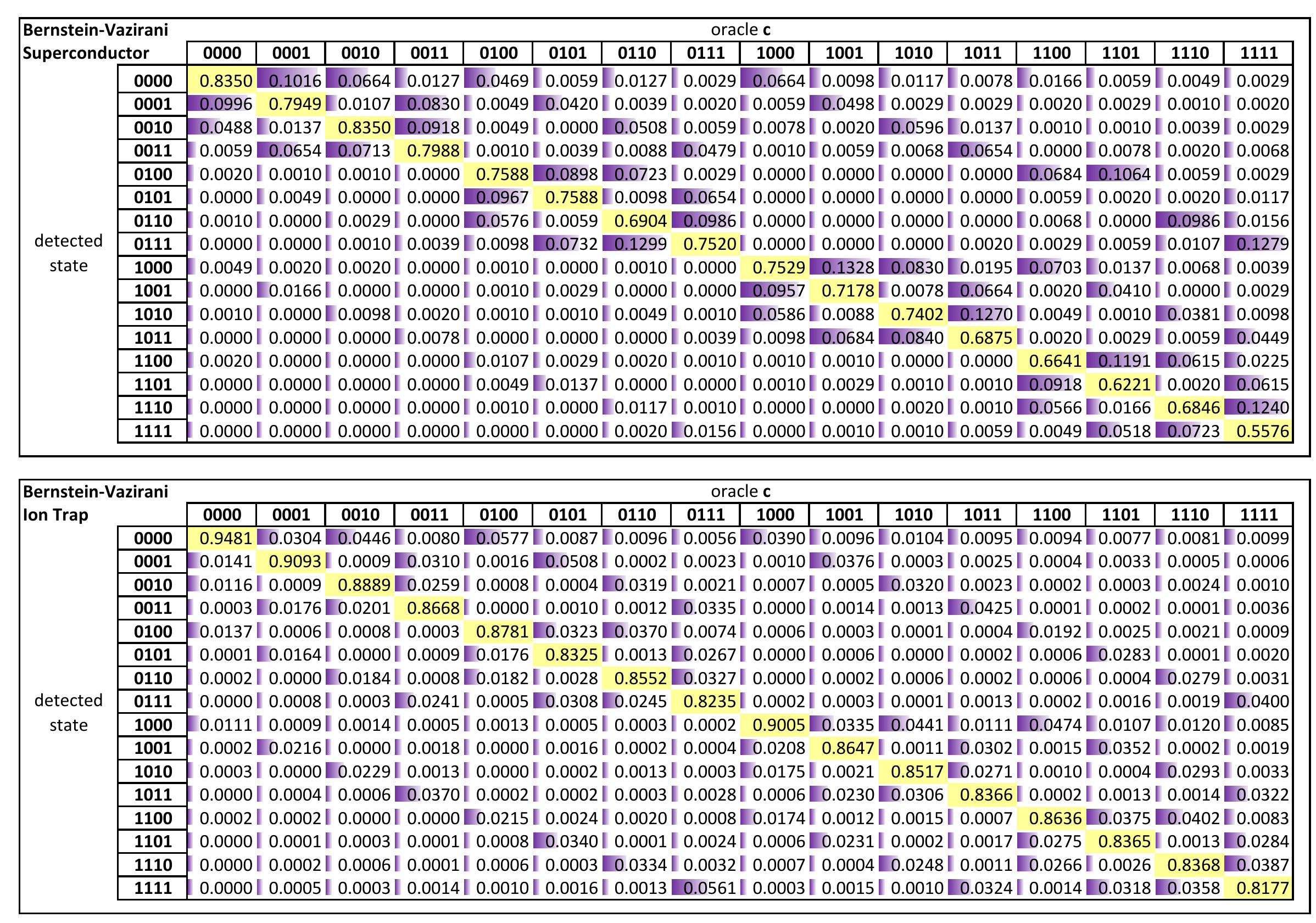}
\caption{Numerical quantum computer 4-qubit input/output matrix for the Berstein-Vazirani algorithm for the superconductor system (top) and ion trap system (bottom), corresponding to Figs. 4(a1) and 4(b1) of the main text.}
\end{figure*}

\begin{figure*}[b]
\includegraphics[height=0.9\textwidth, angle=90]{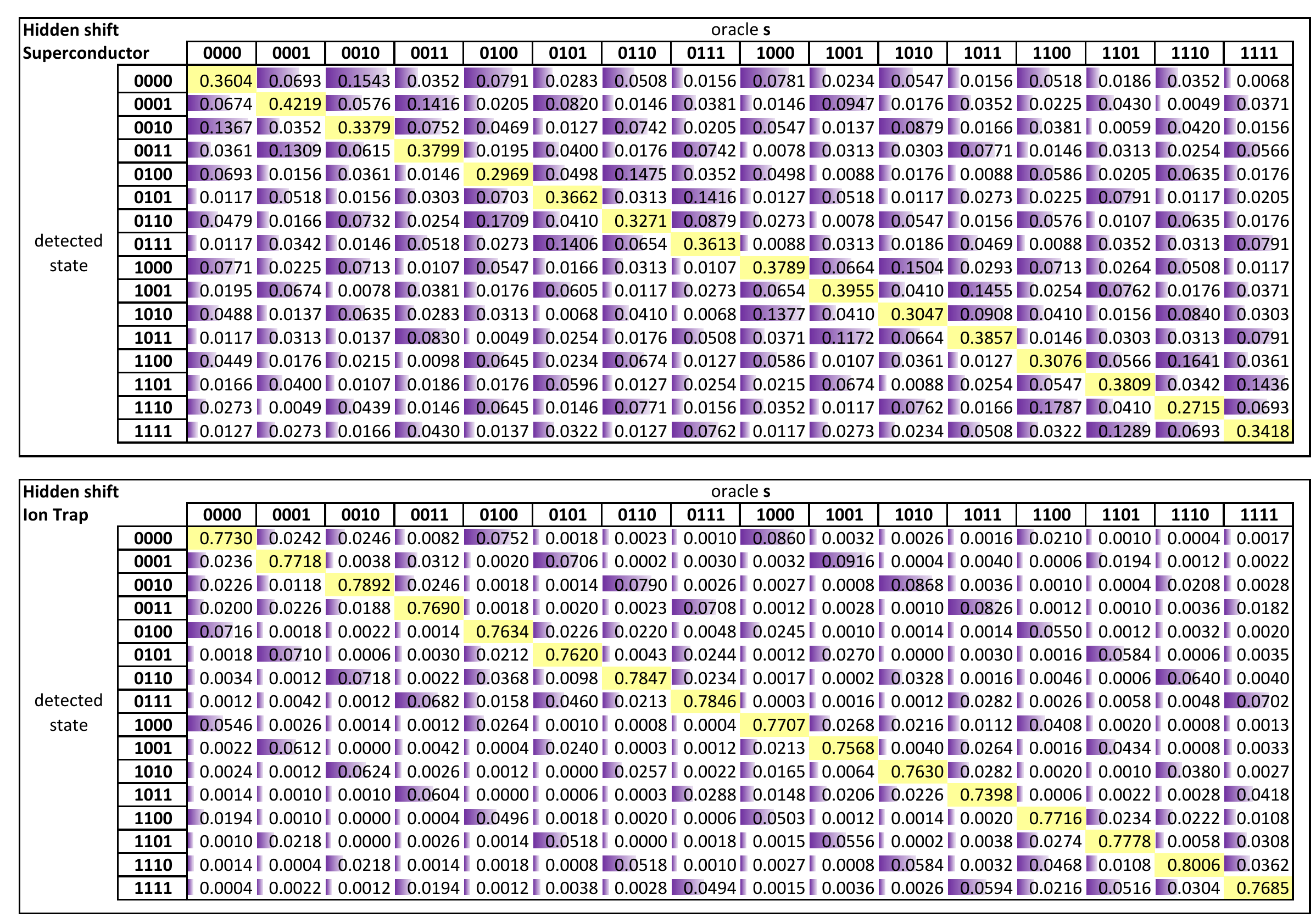}
\caption{Numerical quantum computer 4-qubit input/output matrix for the Hidden Shift algorithm for the superconductor system (top) and ion trap system (bottom), corresponding to Figs. 4(a2) and 4(b3) of the main text.}
\end{figure*}


\begin{thebibliography}{10}

\bibitem{IBM}
(2016).
\newblock IBM Quantum Experience, http://www.research.ibm.com/quantum.

\bibitem{Debnath16}
Debnath S et~al. (2016) Demonstration of a small programmable quantum computer
  module using atomic qubits.
\newblock {\em Nature} 536:63--66.

\bibitem{Barends14}
Barends R et~al. (2014) Superconducting quantum circuits at the surface code
  threshold for fault tolerance.
\newblock {\em Nature} 508:500--503.

\bibitem{Corcoles2015}
C\'{o}rcoles AD et~al. (2015) Demonstration of a quantum error detection code
  using a square lattice of four superconducting qubits.
\newblock {\em Nature Comm.} 6:7979.

\bibitem{Riste2015}
Rist\'{e} D et~al. (2015) Detecting bit-flip errors in a logical qubit using
  stabilizer measurements.
\newblock {\em Nature Comm.} 6:6983.

\bibitem{Ofek16}
Ofek N et~al. (2016) Extending the lifetime of a quantum bit with error
  correction in superconducting circuits.
\newblock {\em Nature}.

\bibitem{Takita16}
Takita M et~al. (2016) Demonstration of weight-four parity measurements in the
  surface code architecture.
\newblock {\em arXiv:1605.01351v2}.

\bibitem{Monz16}
Monz T et~al. (2016) Realization of a scalable shor algorithm.
\newblock {\em Science} 351(6277):1068--1070.

\bibitem{cheung2007translation}
Cheung D, Maslov D, Severini S (2007) Translation techniques between quantum
  circuit architectures in {\em Workshop on Quantum Information Processing}.

\bibitem{Olmschenk07}
Olmschenk S et~al. (2007) Manipulation and detection of a trapped
  ${\mathrm{yb}}^{+}$ hyperfine qubit.
\newblock {\em Phys. Rev. A} 76:052314.

\bibitem{Molmer99}
M\o{}lmer K, S\o{}rensen A (1999) Multiparticle entanglement of hot trapped
  ions.
\newblock {\em Phys. Rev. Lett.} 82:1835--1838.

\bibitem{Solano99}
Solano E, de~Matos~Filho RL, Zagury N (1999) Deterministic bell states and
  measurement of the motional state of two trapped ions.
\newblock {\em Phys. Rev. A} 59:R2539--R2543.

\bibitem{Milburn00}
Milburn G, Schneider S, James D (2000) Ion trap quantum computing with warm
  ions.
\newblock {\em Fortschritte der Physik} 48(9-11):801--810.

\bibitem{Zhu06}
Zhu SL, Monroe C, Duan LM (2006) Trapped ion quantum computation with
  transverse phonon modes.
\newblock {\em Phys. Rev. Lett.} 97:050505.

\bibitem{Choi14}
Choi T et~al. (2014) Optimal quantum control of multimode couplings between
  trapped ion qubits for scalable entanglement.
\newblock {\em Phys. Rev. Lett.} 112:190502.

\bibitem{Nielsen11}
Nielsen MA, Chuang IL (2011) {\em Quantum Computation and Quantum Information:
  10th Anniversary Edition}.
\newblock (Cambridge University Press, New York, NY, USA), 10th edition.

\bibitem{devoret13}
Devoret MH, Schoelkopf RJ (2013) Superconducting circuits for quantum
  information: An outlook.
\newblock {\em Science} 339(6124):1169--1174.

\bibitem{Koch2007}
Koch J et~al. (2007) Charge-insensitive qubit design derived from the cooper
  pair box.
\newblock {\em Phys. Rev. A} 76:042319.

\bibitem{IBMpriv}
\newblock J. Gambetta, J. Chow (2016), private communication.

\bibitem{Geerlings13}
Geerlings K et~al. (2013) Demonstrating a driven reset protocol for a
  superconducting qubit.
\newblock {\em Phys. Rev. Lett.} 110:120501.

\bibitem{Chow10}
Chow JM et~al. (2010) Optimized driving of superconducting artificial atoms for
  improved single-qubit gates.
\newblock {\em Phys. Rev. A} 82:040305.

\bibitem{Chow11}
Chow JM et~al. (2011) Simple all-microwave entangling gate for fixed-frequency
  superconducting qubits.
\newblock {\em Phys. Rev. Lett.} 107:080502.

\bibitem{Maslov16}
Maslov D (2016) Basic circuit compilation techniques for an ion-trap quantum
  machine.
\newblock {\em arXiv:1603.07678v3}.

\bibitem{Amy13}
Amy M, Maslov D, Mosca M, Roetteler M (2013) A meet-in-the-middle algorithm for
  fast synthesis of depth-optimal quantum circuits.
\newblock {\em IEEE Transactions on Computer-Aided Design of Integrated
  Circuits and Systems} 32:818 -- 830.

\bibitem{Zajac16}
Zajac DM, Hazard TM, Mi X, Nielsen E, R. PJ (2016) Scalable gate architecture
  for densely packed semiconductor spin qubits.
\newblock {\em arXiv:1607.07025}.

\bibitem{Barenco95}
Barenco A et~al. (1995) Elementary gates for quantum computation.
\newblock {\em Phys. Rev. A} 52:3457--3467.

\bibitem{Shende09}
Shende VV, Markov IL (2009) On the cnot-cost of toffoli gates.
\newblock {\em Quantum Info. Comput.} 9(5):461--486.

\bibitem{divincenzo98}
DiVincenzo DP (1998) Quantum gates and circuits.
\newblock {\em Proceedings: Mathematical, Physical and Engineering Sciences}
  454(1969):261--276.

\bibitem{Song04}
Song G, Klappenecker A (2004) Optimal realizations of simplified toffoli gates.
\newblock {\em Quantum Info. Comput.} 4(5):361--372.

\bibitem{Bernstein97}
Bernstein E, Vazirani U (1997) Quantum complexity theory.
\newblock {\em SIAM J. Comput.} 26:1411--1473.

\bibitem{Dam02}
van Dam W, Hallgreen S, Lawrence I (2006) Quantum algorithms for some hidden
  shift problems.
\newblock {\em {SIAM} J. Comput.} 36(3):763--778.

\bibitem{Childs13}
Childs A, Kothari R, Ozols M, Roetteler M (2013) {Easy and Hard Functions for
  the Boolean Hidden Shift Problem} in {\em Proc. TQC 2013}, Leibniz
  International Proceedings in Informatics (LIPIcs).
\newblock (Schloss Dagstuhl--Leibniz-Zentrum fuer Informatik), Vol.{}~22, pp.
  50--79.

\bibitem{Roetteler10}
Roetteler M (2010) Quantum algorithms for highly non-linear boolean functions
  in {\em Proceedings of the 21st Annual ACM-SIAM Symposium on Discrete
  Algorithms (SODA’10)}.
\newblock pp. 448--457.

\bibitem{Brecht2016}
Brecht T et~al. (2016) Multilayer Microwave Integrated Quantum Circuits for Scalable Quantum Computing.
\newblock {\em Nature J. Quantum Inf.} DOI: 10.1038/npjqi.2016.2.

\bibitem{Monroe14}
Monroe C et~al. (2014) Large-scale modular quantum-computer architecture with
  atomic memory and photonic interconnects.
\newblock {\em Phys. Rev. A} 89:022317.

\bibitem{Kielpinski02}
Kielpinski D, Monroe C, Wineland DJ (2002) Architecture for a large-scale
  ion-trap quantum computer.
\newblock {\em Nature} 417:709--711.

\end{thebibliography}
\end{document}